\documentstyle[11pt]{article}

\begin{document}

\vspace*{2cm}
\begin{center}
 \Huge\bf
Relation of uncertainty for time 
\vspace*{0.25in}

\large

Alexander K.\ Guts
\vspace*{0.15in}

\normalsize

Department of Mathematics, Omsk State University \\
644077 Omsk-77 RUSSIA
\\
\vspace*{0.5cm}
E-mail: guts@univer.omsk.su

\vspace*{0.5cm}
January 12, 2001\\
\vspace{.5in}
ABSTRACT
\end{center}

We introduce two time: deterministic Newton time-stream $t$ and stochastic
time-epoch $\tau$. The relation of uncertainty for time-epoch of physical 
events
$$
\Delta\tau\Delta D \geq c_1,\eqno(*)
$$
where $c_1=const$, is proved.
The function
$$
D(t)= - c_1\frac{d}{dt}\ln f_\tau (t),
$$
characterizes {\it velocity of disorganization} of the event-phenomena;  
$f_\tau (t)$ is density of probability of time-epoch $\tau$.

The relation (*) is verified not by means of experiment that is traditional 
for physics, but with the help of    the reference to datas of historical 
science.

\vspace{.3in}

\newpage

\setcounter{page}{1}

\def\R{{{\rm I} \! {\rm R}}}
\def\M{{\cal M}}

In the World of events $ \M $ we select such property as the time order.
The time order contacts with such concept as a stream of time.
Events are developed (unwrapped) before the observer consistently, in time.
It means that for measurement of time the special  measuring tool of the 
{\it duration}
of the phenomenon in time and named a watch is used. With the help of watch
 to each
event the concrete number named (time) moment of event or his(its) epoch 
is attributed.
The time order allows to compare epoch of any events.

However the time stream due to which the phenomena consisting of events
are developed (unwrapped) consistently, event behind event, is given to the
person as noted philosopher Kant, a priori, from birth. It is consequence of 
such fact
that the person has topologically trivial 4-dimensional body \cite {guc1,guc2}.
In other words, time  as a stream is only subjective perception
(recognition) of the phenomena of the World of the events inherent to the 
person.

Therefore it is necessary to assume, that time can show itself in our 
human world,
the world of human subjective representations about the World of events,
 absolutely differently than the time order. As a matter of fact it means,
  that time can find out itself as something that can  {\it violate} time
  ordering in deployment of events! Hence events of which the phenomenon 
consists,
  can receive epoches with violation of the time order.

Whether means it, what time can have properties similar to a random variable?
Anyway it is necessary to try to apply principles of probability theory to the
 description of time.

We shall accept further that the choice of the epoches (moments) of time 
which are
attributed to events of the phenomenon with the help of some fixed watch 
can be casual.

Let's forget for simplicity about such concept as a place of event.
In this case events in the World of events can be distinguished only with
 the help of the time order and formally it means that the World of 
events $ \M $
  is the linear ordered continuum like real straight line $ \R $.

Let's assume that we choose watch $t $ which allow each event $x $ to
 attribute the moment of time appropriate to him, i.e. epoch $ \tau $.
 We shall accept that each event gets {\it random epoch}.
  It is understood as the following. So far as event (atomic event or 
the elementary
   phenomenon in the sense of A.D.Alexandrov) is some idealization,
   it should occupy only an instant $ \tau $ in a time-stream $t $.
It is accepted in the theory of a relativity. But actually it
{\it is stretched in a time-stream} $t $ and consequently its epoch
$ \tau $ is absolutely precisely unknown, though must lie on some concrete
segment $ [\tau, \tau + \Delta\tau]$ of time $t $. Hence epoch $ \tau $
of event $x $ is a random variable $ \tau: < X, {\bf S}, {\bf P} > \to \R $, 
where
 $X $ is probability space of events, $ {\bf S} $ \ \  is\ $\sigma$-algebra 
on $X $,
 $ {\bf P} $ is a probability measure on $X $.

Identifying space of events $X $ with the World of events $ \M $, and 
considering
that $ \M $ is  real straight line $ \R $, we receive time-epoch $ \tau (t) $
 as a random variable given in a time-stream $t $.

Event in probability theory is a measurable subset of space $X $.
In our terminology the concept of the phenomenon  corresponds to concept 
of event in
probability theory.
 In turn the events which consist of the phenomenon are elements of set 
$X $ which
 in probability
  theory correspond to elementary events. In terminology of Minkowski 
events are points of
   the World of events $ \M $. But it is obvious that this is  simplification
   accepted in this theory.

So, we shall accept that property of time which is shown in "choice"\
of the moment of time which corresponds to event is a random variable which we
shall name time-epoch.

Let $f_\tau (t) $ be a density of distribution of time-epoch
$ \tau $ satisfying two conditions
$$ {\bf M}\tau =\int\limits_{-\infty}^{+\infty}t f_\tau
(t)dt=0 \eqno(1)
$$
$$
 \lim\limits_{t\to\pm\infty} tf_\tau (t)=0 \eqno(2)
$$
The first condition as it is known is not something important and is
connected to  a choice of origin of time $t$.

Let
$$
D(t)= - c_1\frac{d}{dt}\ln f_\tau (t), \eqno(3)
$$
where $c_1=const$.
We have
$$
{\bf M}D= - c_1\int\limits_{-\infty}^{+\infty}\left(\frac{d}{dt}\ln f_\tau
(t)\right) f_\tau (t) dt = - c_1\int\limits_{-\infty}^{+\infty}
\frac{1}{f_\tau(t)}\frac{df_\tau(t)}{dt}f_\tau (t)dt =
$$
$$
 =-
c_1\int\limits_{-\infty}^{+\infty}df_\tau (t) =
- c_1f_\tau (t)|_{-\infty} ^{+\infty} =0.
$$
Therefore a mean square deviation of $D$
$$
 \Delta D =\sqrt{ {\bf
D}D}=\sqrt{ {\bf M}D^2- ({\bf M}D)^2 }= \sqrt{ {\bf M}D^2}.\eqno(4)
$$
Let's find out sense of  $D $ defined by the formula (3). As $f_\tau (t) $
density of distribution of size $ \tau $ its(her) sense is probability of that
event will receive the epoch laying on a piece of a time-stream $ [t, t+1] $,
where 1 is a standard unit of measurement of time.
Let's find out sense of  $D $ defined the formula (3).
As $f_\tau (t) $ is density of distribution of  $ \tau $, then its sense
is probability of that event will receive the epoch laying on  segment
$ [t, t+1] $ of a time-stream, where 1 is a standard unit of
measurement of time.

But then by analogy to the Boltzmann formula  for entropy, it is possible
to declare
 that $-c_1\ln f_\tau (t) $ is entropy of  time-epoch. In other words, it
 characterizes a measure of disorganization of event as the phenomenon.
Therefore  $D (t) $ characterizes {\it velocity of disorganization}
the event-phenomenon.

As it will be shown below this velocity the is more, than temporal borders
are closer for  localization of the phenomenon in a stream of time. We can
deduce  now some law to which time-epoch satisfies.

\vspace*{0.5cm}

\noindent {\bf Theorem.} {\it If the conditions {\rm (1), (2)}
are held, then  relation of uncertainty

$$
       \Delta\tau\Delta D \geq c_1 \eqno(5)
$$
is true.}

 {\bf Proof.}

To prove (5) we apply the method of Weyl \cite[p.69-70]{Landau}.

We have  the inequality
$$
0\leq \int\limits_{-\infty}^{+\infty}\left(\alpha t \sqrt{f_\tau
(t)} +\frac{d}{dt}\sqrt{f_\tau (t)}\right)^2dt =
$$
$$
=\alpha^2\int\limits_{-\infty}^{+\infty}t^2f_\tau (t)dt + 2\alpha
\int\limits_{-\infty}^{+\infty}t\sqrt{f_\tau (t)}
\frac{d}{dt}\sqrt{f_\tau (t)}dt +
\int\limits_{-\infty}^{+\infty} \left(\frac{d}{dt}
\sqrt{f_\tau (t)}\right)^2dt \eqno(6)
$$
Let's calculate each of integrals in the right part of inequality (6).
First of all, it follows from (1)
$$
 \int\limits_{-\infty}^{+\infty}t^2f_\tau (t)dt ={\bf M\tau^2}=
 {\bf M\tau^2} -({\bf
M\tau})^2 = {\bf D\tau}. \eqno(7)
$$
By using (2.2) we get
$$
2\int\limits_{-\infty}^{+\infty}t\sqrt{f_\tau (t)}
\frac{d\sqrt{f_\tau (t)}}{dt} dt =
\int\limits_{-\infty}^{+\infty}t \frac{d(\sqrt{f_\tau (t)}
\sqrt{f_\tau (t)})}{dt}dt =
\int\limits_{-\infty}^{+\infty}tdf_\tau (t) =
$$
$$
 =tf_\tau (t)|_{-\infty} ^{+\infty} -
\int\limits_{-\infty}^{+\infty}f_\tau (t)dt = -1.\eqno(8)
$$
And, at last, we have for the third integral
$$
\int\limits_{-\infty}^{+\infty}\left(\frac{d}{dt} \sqrt{f_\tau
(t)}\right)^2dt= \int\limits_{-\infty}^{+\infty}\left( \frac{1}
{\sqrt{f_\tau (t)}}
\frac{d\sqrt{f_\tau (t)}}{dt}\right)^2f_\tau (t)dt=
$$
$$
=\int\limits_{-\infty}^{+\infty}\left( \frac{d}{dt}\ln\sqrt{f_\tau (t)}
 \right)^2f_\tau
(t)dt=\frac{1}{4c_1^2} \int\limits_{-\infty}^{+\infty}
\left(c_1 \frac{d}{dt}\ln f_\tau
(t) \right)^2f_\tau (t)dt =
$$
$$
 =\frac{1}{4c_1^2} {\bf M}D^2=\frac{1}{4c_1^2}(\Delta
D)^2. \eqno(9)
$$
Thus from (6)-(9) we have an inequality
$$
\alpha^2
(\Delta\tau)^2 -1 + \frac{1}{4c_1^2}(\Delta D)^2 \geq 0,
$$
which must be true for all $\alpha$. Hence
$$
1- 4(\Delta\tau)^2 \frac{1}{4c_1^2}(\Delta D)^2 \leq 0
$$
or
$$
       \Delta\tau\Delta D \geq c_1.
$$
Theorem is proved.

\vspace*{0.5cm}

The relation of uncertainty (5) was postulated in \cite {guc3,guc4,guc5}
as one of laws of time. The name was given to it: {\it the law
of uncertainty of
 descriptions}. It was formulated on the basis of the analysis
 of the historical
 sources used by historians for descriptions of events of the past.
Thus the relation
  (5) is proved not by means of experiment thai is traditional for physics,
but with the help of
   the reference to datas of historical science.

Though without any doubt after formalization of concepts
$ \Delta \tau, \Delta D $
that it was not made in \cite {guc3,guc4,guc5} it is
possible to speak and
about experimental check of relation (5).

\medskip


Basis for reception of our result was that circumstance that space-time
$ \M $ has  a dual nature which was incorporated by the founder of this
theory Minkowski \cite{Mink}.
  This duality consists of that on the one hand elements of set
$ \M $ are (atomic)
  {\it events}, and by virtue of it $ \M $ carries the name the
World of events, and on
   the other hand $ \M $ is {\it arithmetic} arena on which the
World of events is realized.

This arithmetic arena is necessary for formalization of the World of events
 to attribute to events the coordinates  as the four of real numbers, to
 world lines of the four of real functions etc. As a rule the researcher
  deals with mathematical space-time which we have named arithmetic arena.

However other side of space-time,  the World of events,
 remained in a shadow and was not formalized! At the beginning of
article   we have identified $ \M $ with probability space of events $X $.
Given probability space $X $  is the formalized World of events.

In other words  it is necessary to look at space-time $ \M $ from two
sides, on the one hand  as on set of the elements named elementary
(atomic) events and   to say that {\it event} is any (measurable) subset of set $ \M $
as it is accepted in probability  theory;  on the other hand  as on
coordinate space (arithmetic arena), for example,  $ \R^4 $ which is
used for formalization of concepts of the World of events.

Let's notice, that at the description of deterministic processes and the
phenomena,
 and also at the description of stochastic processes and the phenomena
which are not   touching a nature as of space, and time, we use only
the second point of view on the World of events,
   but the stochastic processes concerning of nature of time
require to distinguish two sides of  the World of events.

Above we used space-time as coordinate space that elementary event
could receive
epoch on "an axis of time".  This epoch  does not lie in "strictly 
allocated place"
according
 to the "instruction" of the time order, but can occupy any place on
 "an axis of time" not especially
 caring of instructions of the mentioned time order. This concerns and
to spatial
  coordinates of events. So with the formula (5) it is possible to
deduce
   similar formulas for mean square deviations of $x-, y-$ and $z-$
   coordinates of events.

\medskip

Let's note one more circumstance. Time as it is found out in this work can be
not only {\it deterministic time-stream} connected with classical
representation  of Newton about time as about duration and, accordingly,
with concept of the time order,   but can be {\it stochastic time-epoch},
having such characteristic as {\it density} probability.

The last sets in the certain sense intensity of display (demonstration) 
of events of the
 phenomenon on a segment of uniformly (current) time-stream. Here it would
  be pertinent to recollect that  N.A.Kozyrev wrote about {\it density
of time} describing its intensity in his articles \cite{Koz}.
    And though in our case the question is stochastic properties of
time nevertheless      it is possible to be surprised the intuition
of Pulkov astronomer.

\end{document}